\documentclass[preprint,12pt]{elsarticle}

\usepackage{amssymb}
\usepackage{float}
\usepackage[keeplastbox]{flushend}
\usepackage{setspace}
\usepackage{appendix}
\usepackage{graphicx}% Include figure files
\usepackage{dcolumn}% Align table columns on decimal point
\usepackage{bm}% bold math
\usepackage{dcolumn}% Align table columns on decimal point
\usepackage{braket}
\usepackage{caption}
\usepackage{subcaption}
\usepackage{float}
\usepackage{caption}

\journal{Journal of Computational Physics}

\begin{document}

\begin{frontmatter}



\title{Conditional Seq2Seq model for the time-dependent two-level system}


\author[inst1]{Bin Yang}

\affiliation[inst1]{organization={Department of Physics\&Astronomy},%Department and Organization
            addressline={Texas A\&M University},
            city={College Station},
            postcode={77840}, 
            state={Texas},
            country={United States}}

\author[inst2]{Mengxi Wu}
\author[inst1,inst3]{Winfried Teizer\corref{cor1}}

\cortext[cor1]{teizer@tamu.edu}

\affiliation[inst2]{organization={Department of Physics\&Astronomy},
 addressline={Louisiana State University},
            city={Baton Rouge},
            postcode={70803}, 
            state={Louisiana},
            country={United States}
 }

\affiliation[inst3]{organization={Department of Materials Science and Engineering},%Department and Organization
            addressline={Texas A\&M University},
           city={College Station},
            postcode={77840}, 
            state={Texas},
            country={United States}}

\begin{abstract}
We apply the deep learning neural network architecture to the two-level system in quantum optics to solve the time-dependent Schr\"odinger equation.  By carefully designing the network structure and tuning parameters, above 90 percent accuracy in super long-term predictions can be achieved in the case of random electric fields, which indicates a promising new method to solve the time-dependent equation for two-level systems. By slightly modifying this network, we think that this method can solve the two- or three-dimensional time-dependent Schr\"odinger equation more  efficiently  than  traditional approaches.
\end{abstract}

%Graphical abstract
% \includegraphics[width=\linewidth]{abstract.png}

%Research highlights
% \begin{highlights}
% \item Introducing a new network architecture by modifying UNet++ with attention gate for single layer graphene (SLG) optical microscopy (OM) image identification.
% \item  Modified UNet++ with attention gate focuses on the interesting area automatically and classifies the image with a pre-trained classifier, increasing the classification accuracy. 
% \item Modified UNet++ with attention gate has the ability to identify the SLG OM image with  less than 100 training images and reaches $>90\%$ accuracy, exceeding existing networks. 
% \end{highlights}


\end{frontmatter}


\section{\label{sec:level1}Introduction}
Recently, due to great successes in image processing and natural language, machine learning and especially deep learning has attracted increased attention in diverse fields besides computer science, including physics, chemistry and healthcare. The ability of deep neural networks to process high dimensional data makes them novel tools in the physics community. This has proven to be an efficient method for many-body physics in condensed matter, e.g. for the classification of topological phases \cite{topological_phase,topological_states}, quantum many-body state generation \cite{many_body,many_body2}, quantum entanglement detection \cite{quantum_entanglement,quantum_entanglement2,entanglement_entropy} and the simulation of quantum systems \cite{phase}. They have also been employed for learning to identify, classify and even discover particles in high energy physics \cite{high_energy}.

In recent work, machine learning methods focused mostly on time-inde-pendent quantum systems. A convolution neural network (CNN) \cite{cnn} or a fully connected network (FCN) is the usual network architecture to solve such problems, due to their strong ability to identify common patterns in large data sets.  However, in many fields of physics, especially in quantum optics, time is also a crucial parameter that has to be considered. In doing so, the underlying rules in quantum systems transit from the time-independent to the time-dependent  Schr\"odinger equation (TDSE), thus showing more complex and interesting behavior. In quantum optics, an important topic is the investigation of different atomic behaviors through the interaction with ultra-short laser pulses \cite{ultra_short_laser}. In condensed matter, non-equilibrium physics is still one of the unsolved frontiers, in which a lot of non-trivial phases of matter are investigated away from equilibrium \cite{non_equilibrium}.

Here we propose a deep neural network architecture, based on long-short term memory (LSTM) \cite{lstm}, to investigate the interaction of two-level atoms with light, one of the basic time-dependent quantum systems. We chose time-dependent two-level systems as our target because they have a simple form, yet contain rich information about the dynamics explaining many physics phenomena. By only training the network with a simple plane light wave, we show that the network can produce super long-term predictions and fully simulate the evolution of the system under the interaction with random light waves. As we show here, well above 90 percent of the tested samples are well predicted and fitted to the truth. This work suggests a new route to treat time-dependent quantum systems and has great potential to be applied in two or three dimensional environments such as the interaction of ultra-short laser pulses with different novel materials, which is difficult to solve by ordinary algorithms.

\section{\label{sec:level2}Time-dependent Schr\"odinger Equation for Two-level systems}
The wave function for a two-level system can be written as \cite{quantum_optics} :
\begin{equation}
    \ket{\psi} = C_{1} \ket{1} + C_{2}\ket{2},
\end{equation}
where  $\ket{1}$ and $\ket{2}$ are two atomic states and $C_{1}$ and $C_{2}$ are the time-dependent amplitudes, we want to calculate. The TDSE reads:
\begin{equation}
    i\ket{\dot{\psi}} = H \ket{\psi},
\end{equation}
where the Hamiltonian can be separated into a time-independent part $H_{0}$
that comes from the two-level energies and a time-dependent part $H_{I}$ that comes from the laser coupling:
\begin{equation}
    H = H_0 + H_I,
\end{equation}
In matrix form, these two parts can be written as
\begin{equation}
    H_0=\left(
\begin{array}{cc}
 \omega _1 & 0 \\
 0 & \omega _2 \\
\end{array}
\right)
\end{equation}
\begin{equation}
    H_I = E(t) D
\end{equation}
\begin{equation}
    D=\left(
\begin{array}{cc}
 0 & \mu  \\
 \mu  & 0 \\
\end{array}
\right),
\end{equation}
where $\omega_1$ and $\omega_2$ are the energies of the two levels, $D$ is the dipole matrix,
$\mu$ is the dipole matrix element and E(t) is the laser field.
The TDSE in matrix form becomes
\begin{equation}
    \frac{i d }{dt}\left(
\begin{array}{c}
 C_1 \\
 C_2 \\
\end{array}
\right)= \left(
\begin{array}{cc}
 \omega _1 & \mu E(t) \\
 \mu E(t) & \omega _2 \\
\end{array}
\right)\left(
\begin{array}{c}
 C_1 \\
 C_2 \\
\end{array}
\right),
\end{equation}
For given two-level parameters ($\omega_1$, $\omega_2$, $\mu$ and the initial value for $C_1$ and $C_2$) and the laser parameter ($E(t)$), we want to solve for $C_1$  and $C_2$ at any time. After solving for the amplitudes $C_1$ and $C_2$, the time-dependent dipole can be calculated as
\begin{equation}
    d(t) = \mu C^*_1C_2 + c.c.
\end{equation}
The above equation can be solved numerically by the splitting operator method \cite{operator_splitting}.

Here, we want to use the deep learning method to solve the two-level system and find the time dependent dipole $d(t)$ in an electric field $E(t)$. The parameters in the two-level system are: $\omega_1, \omega_2, \mu, C_1, C_2$, the electric field is $E(t)$ and the target we want to predict is  $d(t)$. Since the unit can be scaled, we can set the two-level energy difference to 1, and the initial amplitude can be set to the ground state, without loss of generality.

\section{\label{sec:leve3}Network architecture}
Recurrent neural networks (RNNs), as well as their extensions, such as LSTMs and gated recurrent units (GRUs) \cite{grn}, are often employed as the main tools for time series forecasting in deep learning. They achieve outstanding performance in predicting time series because of their ability to incorporate latent information from past input. However, a common RNN architecture cannot predict very long term time series well because of the accumulated errors. As the prediction is going on, it gets more and more unstable and the loss does not converge properly. To solve this problem, we propose a conditional seq2seq model \cite{seq2seq} in our two-level system.

To fully simulate a quantum system’s long term behavior, only 100 initial time points of data are used to predict the next  $1\times10^4$ time points, which is considered a super long term prediction. The neural network we propose is composed of encoder and decoder parts, as shown schematically in Fig. 1. The encoder and decoder parts both consist of 100 LSTM units and each LSTM contains 400 hidden neurons with $tanh$ activation \cite{grn}. In the encoder part, initial input data, $d(t)$ and $E(t)$, are fed into 100 LSTMs separately, and encoder hidden information is generated as output. In the decoder part, the output from the encoder part together with the start of sequence (SOS) will serve as the key to initiate the prediction. The output of each LSTM is fed into a fully connected layer to generate the predicted value. In contrast to the teacher forcing method in the common seq2seq model, which uses true training data for every input in the decoder part, here we use the last predicted $d(t)$ as the next training input. In this way, the accumulated errors are reduced as the training phase and the reference phase are exactly the same. To measure the performance of the network, we use the root mean squared error (RMSE), which is then used in back-propagation.
\begin{equation}
    L = \sqrt{\frac{1}{N}\sum_i (\tilde{d}_i - d_i)^2}
\end{equation}
To train the network, an Adam optimizer \cite{adam} with a learning rate of $1\times10^{-4}$ is used for stochastic gradient descent to minimize root mean square loss.
\begin{figure}
 \captionsetup{singlelinecheck = false, format= hang, justification=raggedright, font=footnotesize, labelsep=space}
 \includegraphics[width=\columnwidth]{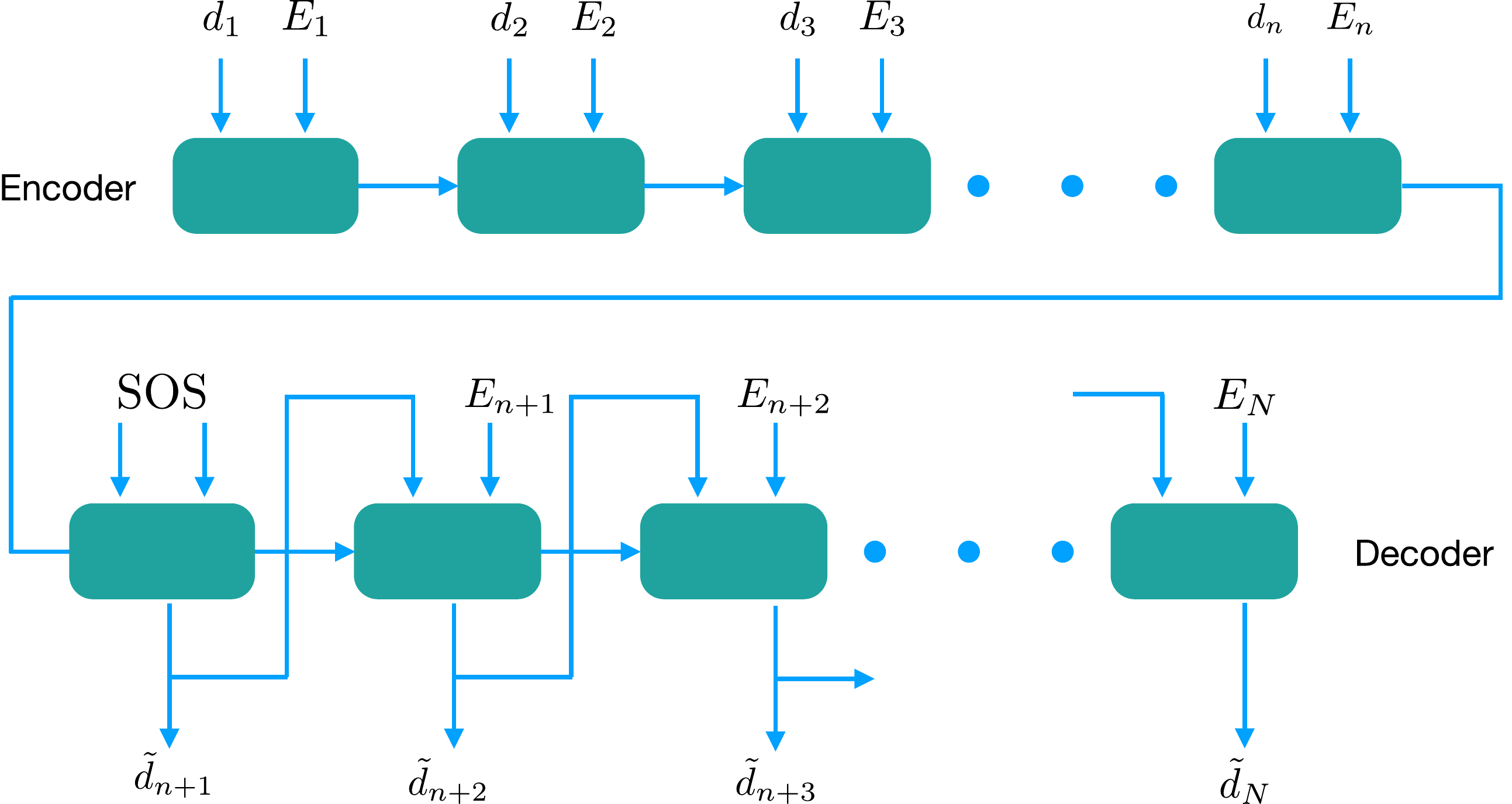}
 \caption{Network architecture of conditional seq2seq model to solve time-dependent Schr\"odinger equation. We use $n=100$ data points to feed the encoder ($d_n$), and predict a total of $N=9900$ data points ($\tilde{d}_{n+1}$ to $\tilde{d}_N$). The new dipole mo-ment time series will be generated continuously by feeding the previous predicted data into encoder with electric field known.}
  \label{fig:lstm}
\end{figure}
\section{\label{sec:level4}Experimental Results}
We first train the network with sinusoidal waves that have a form $E(t)=A\sin(wt)$, since this represents the most basic case in a two level system. In the rotation wave approximation, the analytic solution for two level systems with sinusoidal waves can be obtained. The training samples are calculated by numerical solvers with the splitting operator method[18] and separated by groups. At the first stage, we fix the frequency of the waves at 0.5 Hz, which is half of the atomic frequency between two levels, to avoid the resonance regime. By varying the amplitude from 0.1 to 2.0, about 70 training waves are generated. The reason for setting the maximum amplitude to 2.0 rather than 1.0 is that we find most dipole moment evolution shows nonlinear behavior when the amplitude of the electric field is over 1.0 and it is known from the analytic solution of a two-level system that if the Rabi oscillation amplitude [14] is small, the dipole moment evolution can be expressed by a simple function. To fully train the network to learn the nonlinear physics rules behind the data rather than simple linear dynamics we have to provide enough training samples with amplitude of electric field larger than 1.0. In fact, through the training process, we find that amplitudes of electric field smaller than 0.8 are not significant to the results. By training the samples with electric field waves of amplitudes between 0.8 and 2.0, we can still obtain similar results. By randomly selecting $5\times10^3$ time series with length 200 (100 data for the encoder and 100 for the decoder) in each wave, about $3\times10^5$ training data are generated in which 10 percent will be chosen randomly as validation data. After 100 epochs, the training loss converges to about $1\times10^{-3}$ and the validation loss reaches about $2\times10^{-3}$.

In the test phase, the loss we use is slightly different from the training phase. In the training phase, we use the root mean square loss for each time series piece with length 100 to optimize the parameter: $loss = \sqrt{\frac{1}{N}\sum_{i=1}^{N}(\hat{y_i}-y_i)^2}$, where
N = 100. In the test phase, the test loss is adjusted with respect to the amplitude of electric field. For small amplitude of electric field, the test loss will still be small, even if the percentage loss is $100\%$. Hence, the loss we use for the test is modulated by amplitude of the dipole moment. $Test loss=average rmse/A$, where A represents the amplitude of the true dipole moment. To test the model, we use two different test sets. The first one is a $\sin$ wave function with frequency and amplitude ranging from 0 to 2. In Fig. 2, we show the test loss for changing frequency and amplitude. We find that although the training sample frequency and amplitude are limited, the trained neural network can still make good predictions when $E(t)$ is in the low frequency and amplitude regime. However, when the frequency is above 1 Hz, the dipole prediction starts to be unsatisfactory. To compare the prediction between low and high frequency, we choose two typical prediction images for each of them, which is also shown in Fig. 2(b), (c), (e), (f).  As a second test set we choose the laser pulse waveform $E(t)=A{\sin(wt/20)}^2\sin(wt)$, which is a sin wave with varying amplitude and represents ultra-short pulses in quantum optics. As we can see in Fig. 2(a), (d), similar to the sin wave function, it predicts well at low frequency and amplitude. However, in the regime close to the resonance frequency 1 Hz, the loss is relatively high. As the training sets are only constrained to sin waves with single frequency, the network cannot predict the quantum mechanical rules behind the data very accurately. If the frequency of the test wave is in the resonance regime,  $\Delta=w-\nu$ vanishes, but in the network, $\Delta$ can be expressed as $W_1w-W_2\nu$, where $W_1$ and $W_2$ are both small weights. It can be treated as a fluctuation which can be enlarged through propagation between layers and subsequently can have a large impact on the final output.

\begin{figure*}
 \includegraphics[width=\columnwidth]{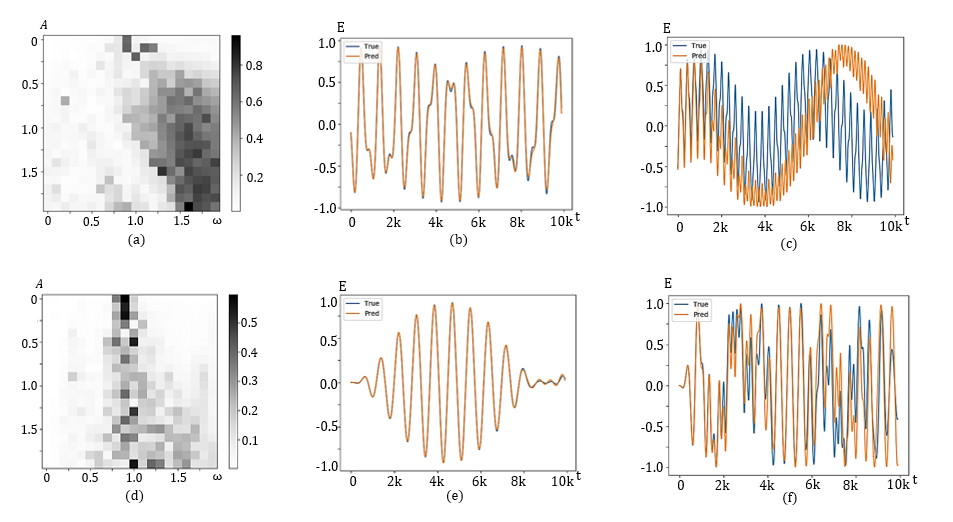}
  \captionsetup{singlelinecheck = false, format= hang, justification=raggedright, font=footnotesize, labelsep=space}
 \caption{(a) Test matrix (x axis represents frequency, y axis represents amplitude and the bar represents mean square loss) loss of model trained on $\sin$ wave with amplitude from 0-2 and frequency 0.5 Hz and tested on $\sin$ wave with both amplitude and frequency ranging from 0 to 2. (b) Prediction of test sample with amplitude 0.2 and frequency 0.5 Hz. (c) Prediction of test sample with amplitude 1.9 and frequency 1.5 Hz. (d) Test matrix loss of same model tested on pulse with both amplitude and frequency ranging from 0 to 2. (e) Prediction of test sample with amplitude 0.2 and frequency 0.5 Hz. (f)Prediction of test sample with amplitude 1.5 and frequency 1.0 Hz.}
  \label{fig:test1}
\end{figure*}
\begin{figure*}
 \includegraphics[width=\columnwidth]{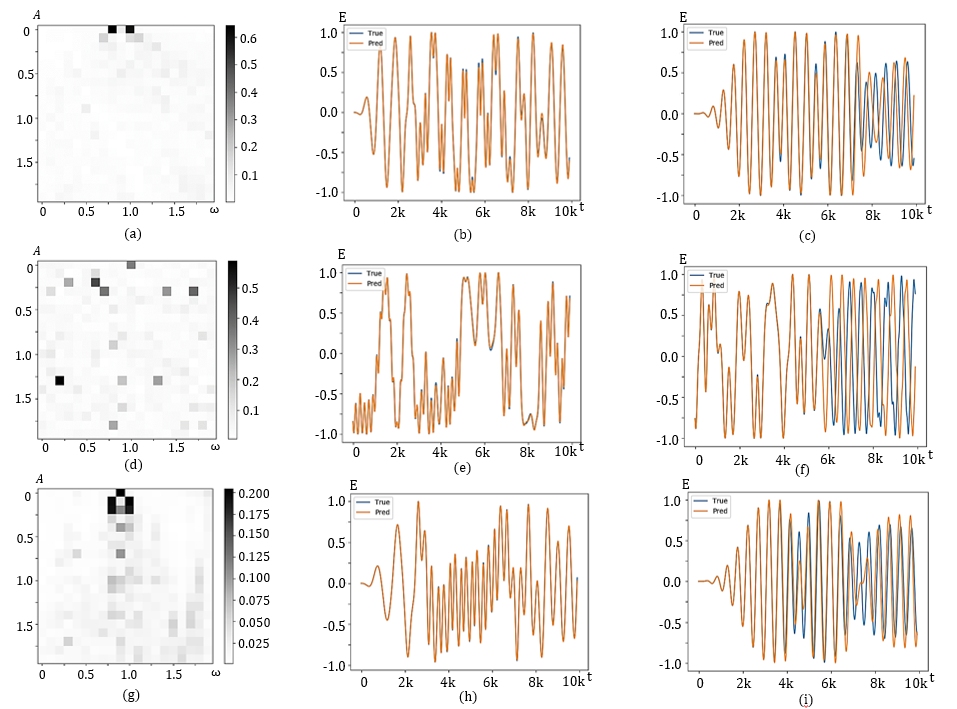}
  \captionsetup{singlelinecheck = false, format= hang, justification=raggedright, font=footnotesize, labelsep=space}
 \caption{(a) Test matrix loss of model trained on sin wave with both amplitude and frequency from 0-2 and tested on regular pulse wave with same amplitude and frequency range. (b) Prediction of test sample with amplitude 1.6 and frequency 0.6 Hz. (c) prediction of test sample with amplitude 0.2 and frequency 1.0 Hz. (d) Test matrix loss of same model tested on random pulse (e)prediction of test sample random selected which show very good prediction. (f) prediction of test sample random selected which deviated to the truth after long time periods.(g) Test matrix loss of same model tested on laser pulse.(h) Prediction of test sample with amplitude and frequency 0.4Hz. (i) Prediction of test sample with amplitude 0.1 and frequency 1.0 Hz.}
  \label{fig:test2}
\end{figure*}
To generate more accurate predictions, we further increase the number of variables and size of training sets by varying both amplitude and frequency from 0.1 to 2. In this way, we can generate 400 different waves and about $2\times10^6$ training data. After 100 epochs, the training loss decreases to $6\times10^{-4}$ and the validation loss decreases to $1\times10^{-3}$ and does not change further in the following epochs.

In order to quantify the performance of this fully-trained model, we first perform a comparison to the test data sets, which contains 400 test samples in the range of $0\leq \omega\leq 2$  and $0\leq E_0\leq 2$ with an offset of 0.038 of the original training data. Fig. 3(a) shows the RMSE of the predicted time-dependent dipole for different $\omega$ and $E_0$. In over 99 percent of the regions, the errors are less than 0.1, with only a few exceptions at very low amplitude. The selected samples are shown in Fig. 3(b), (c). Another test set is based on the pulse electric wave function $E(t)=A{\sin(wt/20)}^2\sin(wt)$ with amplitude and frequency similar to the training sets. We can see in Fig. 3(g) that the predicted data fits very well to the true data, even in the resonant regime, which cannot  be predicted well by previous models. The highest error of prediction is as low as 0.02, which implies very good prediction ability. The selected samples are shown in Fig. 3(h), (i). In order to rigorously test our model further, we expanded our test to include an input wave with random wave form:
\begin{equation}
    E(t) = \mathcal{E}(t) \sin(\omega t),
\end{equation}
where $\mathcal{E}(t)$ is a random function. Fig. 3(e), (f) shows examples of the prediction and Fig. 3(d) shows the loss matrix for 400 random test examples. We can see that the same model, trained on a $\sin$ wave, also predicts very well in the case of a random wave input (above 90 percent accuracy). The overall performance of the model is very impressive since it is only trained on sin electric waves but has the ability to predict dynamics based on random pulses. Based on these extensive tests, we conclude that our deep neural networks are able to predict the time-dependent dipole from the input electric field and generate this prediction from a new unseen data set with amplitude and frequency in a similar range.

To further investigate the model’s capacity, we also test it in the high amplitude and high frequency regime. We start with the $\sin$ plane wave but with large amplitude or high frequency range. As we can see in Fig. 4(a), (d), when the amplitude of the electric field reaches 3 or higher, the loss starts to increase and the predicted dipole does not fit the truth very well after some time periods, since the neural network does not ”see” the data at high amplitude and therefore cannot learn it properly. In the meantime, the frequency of the electric field can reach as high as 4 Hz, which is twice as high as the training sets, while the loss still remains low. When the frequency reaches 5 Hz or higher, the loss starts to diverge and the prediction does not fit the truth very well. The selected samples are shown in Fig. 4(b), (c), (e), (f). This is expected since in such a high frequency regime, there are not enough data points to capture the complex pattern in one time period. For example, when the frequency reaches 6 Hz, there will be about 240 time periods of electric wave in total. Each period only has about 40 points which is not enough to learn the rules behind it.
\begin{figure*}
 \includegraphics[width=\columnwidth]{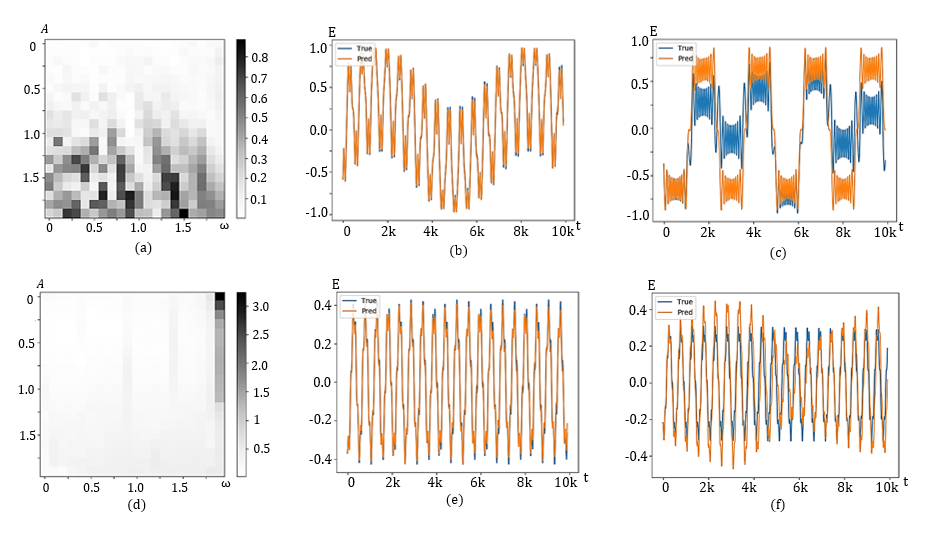}
  \captionsetup{singlelinecheck = false, format= hang, justification=raggedright, font=footnotesize, labelsep=space}
 \caption{(a) Test matrix loss of model trained on sin wave with both amplitude and frequency from 0-2 and tested on sin wave with amplitude from 2-4. (b) Prediction of test sample with amplitude 2.4 and frequency 0.8 Hz. (c) Prediction of test sample with amplitude 3.3 and frequency 0.1 Hz (small deviated to the truth). (d) Test matrix loss of same model tested on pulse with frequency ranging from 4 to 6Hz. (e) Prediction of test sample with amplitude 0.7 and frequency 4.5Hz. (f) Prediction of test sample with amplitude 0.7 and frequency 5.8Hz.}
  \label{fig:test3}
\end{figure*}
\begin{figure*}
 \includegraphics[width=\columnwidth]{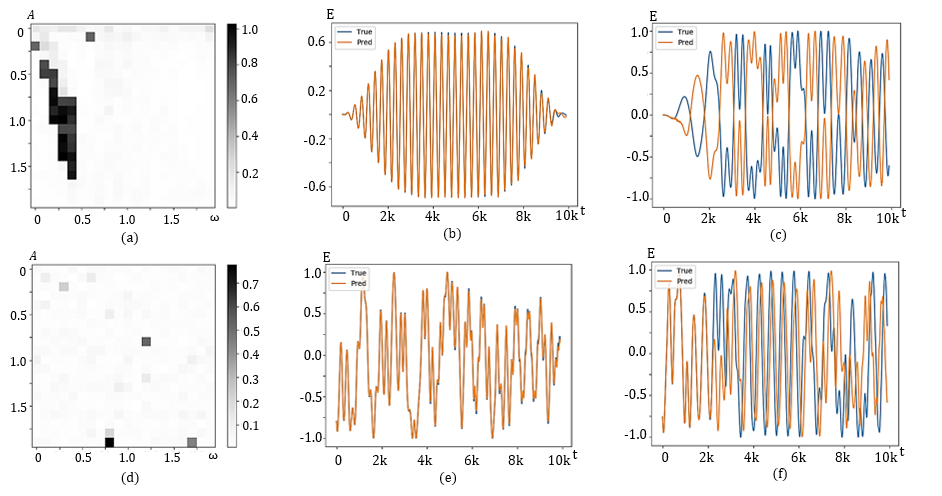}
  \captionsetup{singlelinecheck = false, format= hang, justification=raggedright, font=footnotesize, labelsep=space}
 \caption{(a) Test matrix loss of model trained on random pulse wave and tested on regular pulse wave with same amplitude and frequency range. (b) Prediction of test sample with amplitude 1.6 and frequency 1.6 Hz. (c) Prediction of test sample with amplitude 0.8 and frequency 1.2 Hz. (d) Test matrix loss of same model tested on random pulse. (e) Prediction of test sample random selected which show very good prediction. (f) Prediction of test sample random selected which deviated to the truth after long time periods.}
  \label{fig:test4}
\end{figure*}

To compare with the model trained with sin wave training sets, we also train the neural network with the random pulse training sets. We generate about 400 random pulse waves, as described before, to build up the training sets. After about 100 epochs, both training loss and validation loss drops to $2\times10^{-3}$. In the test phase, we still choose two different test sets like before. One is a regular pulse and another is a random pulse. As we can see in Fig. 5(a), (d), in each case, the overall loss is similar to the model trained with the sin electric waves. In some random points, the loss of the model trained with random pulses is even larger than that trained with sin waves, which indicates that it is enough to predict the new wave by only training with simple sin electric waves with constrained amplitude and frequency. The selected samples are shown in Fig. 5(b), (c), (e), (f).

At last, although it is very common to use ultra-short pulses or plane waves in quantum optics, to prove the generality of our model trained with sinusoidal waves, we also test it with the linear function $E(t)=0.01A_{1}*A_{2}*t$, where $A_{1}$ and $A_{2}$ are two constants ranging from 0 to 2. In Fig. 6(a), we show the testing loss with respect to $A_{1}$ and $A_{2}$ respectively. When the maximum of the linear function is less than 4, the predicted value fits very well to the truth. Only when the maximum reaches 4 or higher, the prediction starts to deviate from the truth and becomes unstable. We can expect this departure from the truth at high amplitude, as we only have training data with an amplitude up to 2 and the network cannot predict data with big differences from the training data. The selected samples are shown in Fig. 6(b), (c).
\begin{figure*}[h]
 \includegraphics[width=\columnwidth]{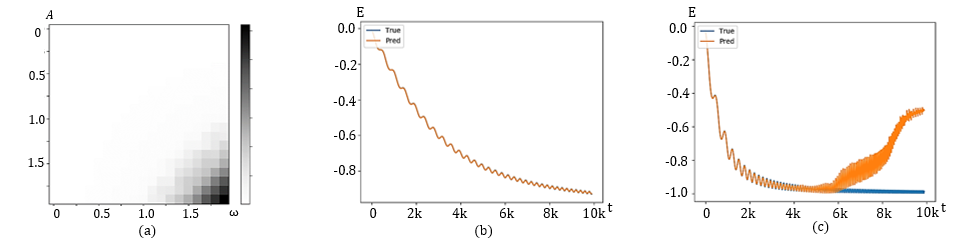}
  \captionsetup{singlelinecheck = false, format= hang, justification=raggedright, font=footnotesize, labelsep=space}
 \caption{(a) Test matrix loss of model trained on sin wave with both amplitude and frequency from 0-2 and tested on linear potential. (b) Prediction of test sample with maximum potential 0.96. (c) Prediction of test sample with maximum potential 3.8.}
  \label{fig:test5}
\end{figure*}

\section{Conclusion}
Although long term prediction by neural networks is hard to achieve due to the long time dependence, we can obtain relatively high accuracy even after 104 time periods in time-dependent two-level systems. We attribute the reason for this to the less complicated quantum linear response system compared to the non-linear one. For the non-linear system, we have to modify the network to contain more hidden states and skip-connections for long time dependence learning. However, the number of parameters for learning is very large and exceeds our computer power. In addition, by replacing the one-dimensional electric field by two or three dimensional image data, we expect our network to be able to solve the two or three dimensional time-dependent equation rather than be limited to the one-dimension case.

\noindent\textbf{Author contributions}

\noindent Bin Yang: Conceptualization, Methodology, Writing - Original Draft, Visualization. 

\noindent Mengxi Wu: Validation, Investigation, Software, Resources.

\noindent Winfried Teizer: Writing - Review $\&$ Editing, Project administration, Supervision. All authors have given approval to the final version of the manuscript.
\\

\noindent\textbf{Declaration of competing interest}

\noindent The authors declare that they have no known competing financial interests or personal relationships that could have appeared to influence the work reported in this paper.
\\

\noindent\textbf{Acknowledgements}

\noindent We gratefully acknowledge financial support from the Center for Nanoscale Science and Technology at Texas A$\&$M University.

 \bibliographystyle{elsarticle-num} 

 \bibliography{myreference}





% \end{thebibliography}
\end{document}